

\magnification\magstep1
\def\ref#1{$^{[#1]}$}
\parskip = 6 pt
\def\pagenumber{\footline={\hss\tenrm\folio\hss}}
\def\half{{1\over 2}}

\def\tu{\tilde u}

\def\tv{\tilde v}

\def\t#1,#2,{\{#1\otimes #2\}}
\def\o{\otimes}

\def\sect#1{\vskip 12pt \leftline{\bf #1} \vskip 1pt}

\def\E-a{E_{-\alpha}}

\input amssym.def    
\input amssym.tex
\def\P{{\Bbb P}}  

\def\tu{\tilde u}
\def\tv{\tilde v}
\def\twoway{\longrightarrow \atop \longleftarrow}
\def\bridgek{\mathop{\twoway}_{{\cal U}_{a-}}^{{\cal U}_{a+}}}
\def\bridgeu{\mathop{\twoway}_{u_{a-}}^{u_{a+}}}

\def\MESS{6}
\def\RESH{7}
\def\KR{8}
\def\AM{9}

\nopagenumbers
\rightline{DAMTP-94-41}
\rightline{hep-th/9407116}
\vskip 20pt
\centerline{\bf QUANTISATION OF A PARTICLE MOVING ON A GROUP MANIFOLD}
\vskip 50 pt

\centerline {\bf  Meifang Chu and Peter Goddard}
\vskip 18pt
\centerline {\it D.A.M.T.P.}
\vskip 1pt
\centerline{\it University of Cambridge}
\vskip 1pt
\centerline{\it Silver Street, Cambridge, CB3 9EW}
\vskip 1pt
\centerline{\it United Kingdom}
\vskip 60pt

\centerline {\bf ABSTRACT}
\vskip 6 pt
{\rightskip=18 true mm \leftskip=18 true mm

The Hilbert space of a free massless particle moving on
a group manifold is studied in details using canonical quantisation.
While the simplest model is invariant under a global symmetry,
$G \times G$, there is a very natural way to
``factorise" the theory so that only one copy of the global
symmetry is preserved. In the case of $G=SU(2)$,
a simple deformation of the quantised theory is proposed to give
a realisation of the quantum group, $U_t(SL(2))$.
The symplectic structures of the corresponding classical theory
is derived. This can be used, in principle, to obtain a
Lagrangian formulation for the $U_t(SL(2))$ symmetry.

}
\vskip 120pt
\centerline{18 July, 1994}

\noindent
\vfill\eject
\pagenumber

\sect{Introduction}

The algebraic structures of quantum groups $U_t(G)$
have been widely studied for some time\ref{1}.
However, its geometrical meaning as a symmetry in quantum field theory
is still not well understood.
It is therefore an intriguing task to obtain a Lagrangian
formulation of the quantum group symmetry.
One interesting possibility is to understand the
quantum group structures in the chirally factorised
two-dimensional Wess-Zumino-Witten models.\ref{2,3}
However, it is very difficult to obtain an explicit formulation
of the quantum group structures in these chiral models
because their defining fields, so-called chiral
vertex operators, are non-local operators.
It turns out that we can study a simpler model
which describes a particle moving on the manifold of a Lie group $G$.
The idea is to study whether one can implement the quantum group
structures, viewed as some deformation of $G$, by introducing
nontrivial interaction to the particles.
In other words, one should study a Lagrangian formulation
of interacting particles with $U_t(G)$ structure where
the deformation parameter $t$ can be related to the coupling constants.
The aim of this paper is to investigate this possibility.

One of the common features for particles and strings moving
on a group manifold is the quadratic Poisson structures
among their defining fields, $U \in G$. In terms of the tensor notation,
$U_1 = U \o 1$ and $U_2=1\o U$, they can be written as
$$
\{U_1, U_2 \}=U_1 U_2 r_{12}.
\eqno(1)
$$
In general, the matrix $r_{12}$ may depend on dynamical variables.
Faddeev et al\ref{3,4} suggested that these Poisson brackets
can be quantised by an exchange relation,
$$
U_1 U_2 = U_2 U_1 B_{12},
\eqno(2)
$$
where the classical limit of the braiding matrix $B$ is
$r_{12}$ in (1).
For a particle moving on $G=SU(2)$, this quantisation was
first discussed in [4], but the Hilbert space structure of
the quantised theory was not fully discussed.

In this letter, we give a systematic derivation of
the canonical structure of a free particle moving on a
compact simple Lie group, $G$.
In order to relate to the zero modes of the WZW model
associated with $G$, we have used the same approach in [2]
to quantise the particle case.
It may not seem natural from the usual quantisation's
point of view, but we will justify this particular approach and
show that there are several advantages.
In particular, it leads to a natural way of factorising
the left-invariant and the right-invariant theory.
Also, it suggests a simple deformation of the non-interacting
model to give a quantum realisation of $U_t(G)$.
This deformation is explained in details
for $U_t(SL(2))$ using two pairs of harmonic oscillators.
We also derive the symplectic two-form which,
in principle, should allow us to write down the
Lagrangian whose symmetry is governed by $U_t(SL(2))$.

\sect{Massless Free Particle Moving on a Group Manifold}

Let $g(\tau)$ denote a particle at time $\tau$ taking values
on a compact simple Lie group $G$ of rank $r$.
Denote $t^a\equiv \{ H^i, E^\alpha \}$, for $i=1,2,..r$
and all the roots $\alpha$, to be a basis of the generators
for the corresponding algebra ${\cal G}$.
These generators are normalised such that $Tr(t^a t^b)=\delta^{ab}$.
Assuming that there is no interaction
among the particles, we can write down the Lagrangian for the
free massless particle moving on $G$ as
$$
L= - {1\over 2} Tr (g^{-1} {\dot g} g^{-1} {\dot g}).
\eqno(3)
$$
This Lagrangian is invariant under a global symmetry $G\times G$,
i.e. $g(\tau) \rightarrow g_1 \cdot g(\tau) \cdot g_2$,
$\forall g_1 \in G, g_2 \in G$.
Consequently, the equations of motion give the conservation of
the two currents that generate the group transformations,
$$
\partial_\tau (g^{-1} {\dot g})=0, \qquad
\partial_\tau ({\dot g} g^{-1})=0.
\eqno(4)
$$

A general solution of these equations can be written as
$g(\tau)=u_0 e^{iP\tau}v_0$ where $u_0$ and $v_0$ are the
constant elements in $G$ given by the initial conditions of (4).
However, this parameterisation is ambiguous because we get the
same solution after redefining the variables by
$u_0\rightarrow u_0 f_1$, $v_0\rightarrow f_2^{-1} u_0 $, and
$e^{iP\tau}\rightarrow f_1^{-1} e^{iP\tau} f_2$ with constant
group elements $f_1,f_2 \in G$.
In order to minimise this ambiguity, we will restrict
$e^{iP\tau}$ to be in the maximal torus $T$ and write the solution as
$$
g(\tau)={\tilde u} e^{iq\cdot H} e^{ip\cdot H\tau} {\tilde v},
\qquad {\tilde u} \in G/H, \qquad {\tilde v} \in H\backslash G.
\eqno(5)
$$
This way, the ambiguity in parameterising the solution in (5)
is fixed up to the Weyl group $W$ and the maximal torus $T$.
In this sense, (5) gives a map between the classical
phase space, $G\times G$, and the configuration space, $G$.

According to Zuckerman\ref{5}, there is a
closed symplectic two-form given by the Lagrangian $L$,
$$
\omega={1\over 2} {\rm Tr} \left\{ \delta g \delta ({{\partial L}\over
{\partial {\dot g}}}) \right\}.
\eqno(6)
$$
This symplectic form determines the Poisson structures of the
theory according to,
$$
\omega = \sum_{i,j} \omega^{ij} \delta A_i \delta A_j,
\qquad \{ f_1, f_2 \}= \sum_{i,j} \omega^{-1}_{ij}
{\delta f_1 \over \delta A_i} {\delta f_2 \over \delta A_j},
\eqno(7)
$$
where $\{ A_i \}$ denote a basis of the coordinates on the
phase space. Since it is the space of all the solutions of
the equations of motion, we substitute the solution in (5) into (6).
The symplectic form is now invertible and it gives the
following Poisson brackets,
$$\eqalign{
& \{q^i,p^j\} = \delta^{ij}, \qquad \{\tu_1,\tv_2 \}=0,
\qquad \{ \tu, p^j \} =0, \qquad \{ \tu, p^j \} =0, \cr
& \{ \tu_1, \tu_2 \} = \tu_1 \tu_2 r_{12} - \tu_2 \{ \tu_1, iq\cdot H_2 \}
-\tu_1 \{ iq\cdot H_1, \tu_2 \}, \cr
& \{ \tv_1, \tv_2 \} = -r_{12} \tv_1 \tv_2  - \{ \tv_1, iq\cdot H_2 \} \tv_2
- \{ iq\cdot H_1, \tv_2 \} \tv_1. \cr}
\eqno(8)
$$
The last two brackets in (8) depend on the Poisson brackets
of $\tu$, $\tv$ with $q \cdot H$.
Their explicit forms depend on the parameterisation of the coset.
However, we do not need to specify them for the purpose of this paper.
In fact, we can simplify the Poisson brackets in (8)
in terms of the new variables, $u\equiv \tu e^{iq\cdot H}$ and
$v\equiv e^{iq\cdot H} \tv$,
$$\eqalign{
& \{ u_1, u_2 \} = u_1 u_2 r_{12}, \cr
& \{ v_1, v_2 \} = -r_{12} v_1 v_2, \cr} \qquad {\rm with} \qquad
r_{12} = \sum_{\alpha \in \Phi} {i\over p\cdot\alpha} E_{\alpha}
\o E_{-\alpha}.
\eqno(9)
$$
We have denoted $\Phi$ to be the space of roots for the Lie algebra
{\cal G}. One can verify that these Poisson brackets
satisfy the Jacobi identities.
It is also straight-forward to check that
$u$ and $v$ transform as the left-covariant and
the right-covariant group elements under the currents $L$ and $R$.
$$\eqalign{
& L\equiv \partial_\tau g \cdot  g^{-1},\qquad \{ L_1, u_2 \} =
{\cal C}_{12} u_2, \cr
& R\equiv g^{-1} \cdot \partial_\tau g,\qquad \{ R_1, v_2 \} =
v_2 {\cal C}_{12} , \cr} \qquad {\rm with}\qquad
{\cal C}_{12}\equiv \sum_{a} t^a \o t_a.
\eqno(10)
$$

Now, we can proceed the quantisation of these Poisson brackets.
The first four brackets in (8) can be replaced
with the following Dirac commutators,
$$
[q^i, p^j ]= i \hbar \delta^{ij}, \qquad [\tu_1, \tv_2 ]=0,
\qquad [ \tu , p^j ] =0, \qquad [ \tv , p^j ] =0.
\eqno(11)
$$
As mentioned in the introduction, we quantise
the quadratic brackets in (9) by the exchange relations,
$$
u_1 u_2 = u_2 u_1 B_{12}, \qquad v_1 v_2 = B^{-1}_{12} v_2 v_1.
\eqno(12)
$$
Quantum consistency requires that the braiding matrix $B$
has to satisfy the following conditions,
$$\eqalignno{
& \hbox{{\it Classical limit}}, \hbox{\hskip 4cm}
B={\cal I}\o {\cal I}+ i\hbar r_{12} + O(\hbar^{-2}), & (13a) \cr
& \hbox{{\it Antisymmetry and Unitarity}}, \hbox{\hskip 1.8cm}
B^{-1} = B^{\dagger}= \P B \P , & (13b) \cr
& \hbox{{\it Jacobi identities}}, \hbox{\hskip 2cm}
B_{23}(p_1)B_{13}(p)B_{12}(p_3)=B_{12}(p)B_{13}(p_2)B_{23}(p),
& (13c) \cr
& \hbox{{\it Locality condition}}, \hbox{\hskip 3.5cm}
[B, e^{iq\cdot(H_1+H_2)} ]=0, & (13d) \cr}
$$
where the subscript of $p$ denotes a shift, e.g.
${\vec p}_2= {\vec p}+ \hbar {\vec H}_2$.
This is because $u$ and $v$ defined in (9) do not commute
with the momentum $p$:
$$
[{\vec p},u_2]=\hbar u_2 {\vec H}_2,\qquad
[{\vec p},v_2]=\hbar {\vec H}_2 v_2.
\eqno(14)
$$
The locality condition in (13d) is required
to ensure that $g(\tau)$ is local in the sense that
their equal-time commutator vanishes,
i.e. $[g_1(\tau),g_2(\tau)]=0$.

The explicit solution for the braiding matrix in (13) can be obtained
in a chosen representation for a given group.
In this paper, we give the solution in the fundamental
representation of $G=SU(N)$. It has the form of
exponentiating the classical r-matrix in (9):
$$
B=\exp \left( -\sum_{\alpha \in \Phi} \theta_{\alpha} E_{\alpha}
\o E_{-\alpha} \right).
\eqno(15)
$$
The coefficients $\theta_\alpha$'s can be determined from
the Jacobi identities in (13c),
$$
\sin \theta_{\alpha} = \hbar / p \cdot \alpha.
\eqno(16)
$$
Although the braiding matrix $B$ depends on the momentum
$p$, the eigenvalues of $\P B$ do not. They are 1 and -1 with
multiplicity ${1\over 2}N(N+1)$ and ${1\over 2}N(N-1)$ respectively.
For $N=2$, this braiding matrix has already appeared in [3].

In order to justify that (11), (12) and (15) determine all
the quantum relations, we show that the current
algebras can be derived from them.
First, the quantum corrections to the $SU(N)$ currents
can be obtained from normal-ordering the classical
currents in (10), e.g. putting $p$ to the right-hand side of $q$.
We can write these currents in terms of $u$ and $v$ respectively:
$$
L= i u \left( p \cdot H \right) u^{-1}+ 2i\hbar H^2,
\qquad R= i v^{-1} \left( p\cdot H \right) v + 2i\hbar H^2.
\eqno(17)
$$
Then, using the braiding matrix in (15), we find that
the transformation laws of the covariant vertex operators are given by
$$
[L_1,u_2]=i\hbar {\cal C}_{12} u_2, \qquad
[R_1,v_2]=i \hbar v_2 {\cal C}_{12}.
\eqno(18)
$$
Following from (17) and (18), the symmetry algebras can be
written in terms of the tensor-Casimir $C_{12}$ defined in (10)
as follows.
$$
[L_1,L_2]=i\hbar[C_{12}, L_2],\qquad [R_1,R_2]=i\hbar[C_{12}, R_2].
\eqno(19)
$$

Furthermore, the quadratic Casimir operators for $L$ and $R$ turn out
to be equal and they depend only on the momentum $p$,
$$
C\equiv - Tr (L^2)\equiv - Tr (R^2)= p^2 - \hbar^2 Tr(H^2 H^2).
\eqno(20)
$$
This implies that the Hilbert space can be decomposed into a
direct sum of the diagonal products of the
irreducible representations of the left and the right
symmetry algebras,
$$
{\cal H}=\bigoplus_\lambda {\cal V}_\lambda \otimes
{\overline {\cal V}}_\lambda .
\eqno(21)
$$

In order to illustrate the factorisation of the Hilbert
space and to study the intertwining operators on this space,
we now give an example of the above quantisation for $G=SU(2)$
in details. In particular,
we will construct the left-covariant vertex $u$ explicitly
in terms of harmonic oscillators and
determine the Hilbert space which it acts on.
To do so, it is very convenient to use the
Euler parameterisation for the solution in (5):
$$
u = e^{ i \half A \sigma_3} e^{i \half D' \sigma_2}
e^{i\half q \sigma_3}, \qquad
v = e^{i\half q \sigma_3}e^{i \half {\overline D}'\sigma_2}
e^{i\half {\overline A} \sigma_3}.
\eqno(22)
$$
The symplectic form in (6) can now be easily inverted
and the quantisation of the Poisson brackets gives
three pairs of harmonic oscillators,
$$
[q, p]=i\hbar, \qquad [A, D]=i\hbar,
\qquad [{\overline A}, {\overline D}]=i\hbar,
\eqno(23)
$$
where we have defined $D \equiv p \cos D'$ and
${\overline D} \equiv p \cos {\overline D}'$.
These harmonic oscillators correspond to the
six degrees of freedom in the phase space.

The left-$SU(2)$ and the right-$SU(2)$ currents in (17)
can now be expressed in terms of the harmonic oscillators as
$$
\eqalign{
L^3= D, &\qquad
L^{\pm}= e^{\pm i A} \sqrt{ (p+D\pm {\hbar \over 2})
(p-D\mp {\hbar \over 2}) }, \cr
R^{3} = {\overline D}, &\qquad
R^{\pm}= e^{\pm i {\overline A}} \sqrt{ (p+{\overline D}\pm
{\hbar \over 2}) (p-{\overline D} \mp {\hbar \over 2})},
\cr}
\eqno(24)
$$
and the quadratic Casimir operators depend only
on the momentum,
$$
C=-Tr(L^2)=-Tr(R^2)=(p + {\hbar\over 2})(p - {\hbar\over 2}).
\eqno(25)
$$
Therefore, the irreducible representations of the
symmetry algebras are labelled by the eigenvalues of
the momentum, $p=(\ell+{1\over 2})$, with the spin
$\ell=0,{1\over 2}, 1,...\infty$. Let us denote the
$(2\ell+1)$-dimensional irreducible representation of
the left-$SU(2)$ by ${\cal V}_\ell$ and the one for
the right-$SU(2)$ by  ${\bar{\cal V}}_\ell$. Then,
the Hilbert space can be decomposed into
$$
{\cal H} =  {\bigoplus}_{\ell=0,{1\over 2}, 1,...\infty}
{\cal V}_\ell \o {\overline {\cal V}}_\ell,
\eqno(26)
$$

One can obtain the states in the Hilbert space as follows.
Define the following orthonormal basis $ \{ |\ell;m,n \rangle \}$
with
$$
|\ell;m,n \rangle \equiv e^{i(\ell+{1\over 2}) q} e^{imA}
e^{in{\overline A}} |0 \rangle.
\eqno(27)
$$
The highest weight state of spin $\ell$ is
$| \ell;\ell,\ell \rangle$ since it is
annihilated by the raising operators $L^+$ and $R^+$.
The rest of the states in the representation can
be obtained by applying the lowering operator
$L^-$ and $R^-$ to the highest weight state:
$$
{\cal V}_\ell \o {\overline {\cal V}}_\ell \equiv
\left\{ L_-^{\ell - m} R_-^{\ell- n}
|\ell;\ell,\ell\rangle, \quad m,n = -\ell,-\ell+1,..\ell \right\}.
\eqno(28)
$$
These current generators act on the states according to
$$
\eqalign{
&L^{\pm}|\ell;m,n \rangle =\sqrt{(\ell\mp m)(\ell\pm m +1)}
|\ell;m\pm 1, n \rangle ,\cr
&R^{\pm}|\ell;m,n \rangle =\sqrt{(\ell\mp n)(\ell\pm n +1)}
|\ell;m, n\pm 1 \rangle.\cr}
\eqno(29)
$$

This realisation makes it easy for us to study the action of
the group element $g(\tau)$ and the left-covariant vertex operator
$u$ on the Hilbert space. For example, in the spin ${1\over 2}$
representation, $g(\tau)$ defined in (5) is a unitary
$2 \times 2$ matrix (normal-ordered),
$$
g_{ab} (\tau) =  a g^+_{ab} - b g^-_{ab}, \qquad {\rm for}
\quad a,b = \pm 1,
\eqno(30)
$$
where $g^{\pm}$ takes a state of spin $\ell$ into a new state
of spin $\ell \pm \half$. To be more precise, we have
$$
g^{\pm}_{ab}(\tau) | \ell;m,n \rangle = \half
e^{\pm i ({\ell+ \half})\hbar \tau} \sqrt{1\pm {nb +1 \over \ell+\half}}
\sqrt{ 1 \pm {a m \over \ell +\half (1\pm 1)} }
\Bigl\vert \ell\pm {1\over 2}; m+ {a \over 2}, n + {b \over 2}
\Big\rangle.
\eqno(31)
$$

On the other hand, applying the left-covariant vertex $u$ on a state
$|\ell;m,n \rangle$ in ${\cal H}$, we find that it gives
the Wigner coefficients\ref{\MESS} of $SU(2)$ according to
$$
u_{ab} |\ell;m,n \rangle = n_{ab}
(-1)^{{b\over 4}(1+a)-1-m-\ell} \sqrt{2\ell +1}
\left( \matrix {\ell & \half \cr m & {a\over 2} \cr}
\matrix {\ell+{b\over 2} \cr -m -{a\over 2} \cr} \right)
\Bigl\vert \ell+ {b\over 2}; m+ {a\over 2},n \Big\rangle
\qquad \not\in {\cal H},
\eqno(32)
$$
where $n_{ab}=1$ except for $n_{-+}=-1$.
Unlike $g(\tau)$, $u$ takes a state in ${\cal H}$ outside of
${\cal H}$. This is because the spin of the representation
is changed while the quantum number $n$ does not.
If we omit $n$ by projecting it down to $0$, then $u$
becomes a well-defined operator in the left-projected space,
$$
{\cal P}_L({\cal H}) \equiv \bigoplus_{\ell\ge 0} {\cal V}^L_\ell,
\quad {\rm where} \quad {\cal V}^L_\ell \equiv
\left\{ L_-^{\ell -m} |\ell;\ell,0 \rangle, \quad
m= -\ell,-\ell+1,..\ell. \right\}.
\eqno(33)
$$
This projected space can be identified as
the Hilbert space for the left-invariant theory
defined by $u$. All the states in different representations
are connected to one another by $u$ in the following way:
$$
{\cal V}^L_0 \qquad \bridgeu \qquad {\cal V}^L_{1\over 2} \qquad
\bridgeu \qquad {\cal V}^L_1 \qquad \bridgeu \cdot \cdot \cdot \cdot
\eqno(34)
$$
Notice that the left-index of $u$ labels the new representation and
the right-index labels the weight of the new states.

Similarly, for the right-invariant theory,
$v$ acts on the right-projected Hilbert space with $m=0$,
$$
{\cal P}_R ({\cal H}) \equiv \bigoplus_{\ell\ge 0}
{\overline {\cal V}}^R_\ell, \quad {\rm where} \quad
{\overline {\cal V}}^R_\ell \equiv \left\{ R_-^{\ell -n}
|\ell;0,\ell\rangle, \quad n= -\ell,-\ell+1,..\ell. \right\}.
\eqno(35)
$$
The difference from $u$ is that the right-index of $v$ labels the new
representation and the left-index labels the weight of
the new states.

Another remark is that the exchange matrix $B(p)$ can be
identified as the Racah (6j-symbols) matrix for $SU(2)$
when it acts on a physical state with eigenvalue,
$p=\ell +\half$ of $\ell >1$.
$$
B_{mn,m'n'} (p)= \delta_{m+n,m'+n'}  \left\{ \matrix{
\half & \ell & \ell-{n' \over 2} \cr
\half & \ell - {m+n \over 2} & \ell - {m\over 2} \cr} \right\}
\qquad {\rm for}\quad m,n=\pm 1.
\eqno(36)
$$
Although $B$ is singular when $p$ is at the origin,
i.e. $\ell =- \half$,
the braiding matrix $B(p)$ is well defined in ${\cal H}$
since this representation is excluded in ${\cal H}$,

\sect{Quantum realisation of $U_t(SL(2))$}

In the previous section, we formulated the quantum theory of
the left-$SU(2)$ invariant particle in terms of two pairs of
harmonic oscillators. In this section, we propose a deformation of those
results to give a realisation for $U_t(SL(2))$.
We shall use $t=q^\half$ to be the deformation parameter
rather than $q$ as in [{\RESH}]. This is because
the defining relations for the quantum group actually
depend on $q^\half$ rather than $q$.

Let us recall the defining relations of the generators for
$U_t(SL(2))$,
$$
[H, X_\pm]=\pm X_\pm, \qquad
[X_+,X_-]=[2H], \qquad {\rm with}\quad
[S] \equiv {t^S - t^{-S} \over t-t^{-1}}.
\eqno(37)
$$
The key observation to obtain a realisation of (37)
is that we can modify (deform) the $L^\pm$ generators given in (24)
by replacing the momentum operators $p$ with $[p]$, i.e.
$$
S_3 \equiv D, \qquad
S_\pm \equiv e^{\pm i A}\sqrt{ [p+D\pm {\hbar\over 2}]
[p-D\mp {\hbar\over 2}]}.
\eqno(38)
$$
Using $[q,p]=[A,D]=i\hbar$, we can check that
$$
[S_3,S_\pm]=\pm \hbar S_\pm, \qquad [S_+, S_-]=[\hbar] [2S_3].
\eqno(39)
$$
At the classical limit when $\hbar \rightarrow 0$,
this algebra is not the usual Lie algebra $SL(2)$.
In other words, the deformation occurs already at the classical level.
For $\hbar=1$, (39) gives a realisation of (37) and we shall
set $\hbar=1$ from now on.
The quadratic Casimir operator depends on $p$ only,
$$
C \equiv S_+ S_- + [S_3] [S_3 + 1]
=[p-\half] [p+\half].
\eqno(40)
$$
Thus, the representations of this algebra
are labelled by the eigenvalues of $[p-\half] =[\ell]$.
These representations are irreducible when $\ell$ is
$0,\half,1,{3\over 2}...$
Using the following orthonormal basis,
$$
\left\{ |\ell;m \rangle \equiv e^{i(\ell+\half)q} e^{imA}
|0 \rangle \right\} .
\eqno(41)
$$
we obtain all the states in the spin $\ell$ representation
by applying the lowering operator $S_-$ to the highest weight states,
$|\ell;\ell \rangle $,
$$
{\cal V}_\ell = \left\{ S_-^{\ell-m}|\ell;\ell>, \quad
 m=-\ell,-\ell+1,...,\ell \right\}.
\eqno(42)
$$
While, the highest weight state $|\ell;\ell \rangle $ is
annihilated by the raising operators $S_+$.

Applying the same deformation to the momenta in (22), we obtain the
the following $2\times 2$ matrix, ${\cal U}$,
$$
{\cal U} \equiv e^{{i\over 2}A\sigma_3} \left( \matrix{x&-y \cr
y&x\cr} \right) e^{{i\over 2}q\sigma_3},\qquad \cases{
x\equiv t^{-(D-p)/2} \sqrt{ [p+D]/[2p]}, \cr
y\equiv t^{-(D+p)/2} \sqrt{ [p-D]/[2p]}. \cr}
\eqno(43)
$$
The phase $t^{-(D-p)/2}$ in $x$ and a similar one in $y$
are introduced such that $\det ({\cal U})$ is 1.
Notice that ${\cal U}$ is not unitary any more under this deformation.
We will call ${\cal U}$ the ``Wigner" operator for $U_t(SL(2))$
because its matrix elements between states in ${\cal V}_\ell$ and
${\cal V}_{\ell\pm \half}$ given the Wigner coefficients for
$U_t(SL(2))$. Thus, we have the following intertwining structure,
$$
{\cal V}_0 \qquad \bridgek \qquad {\cal V}_{1\over 2}\qquad \bridgek
\qquad {\cal V}_1  \dots \qquad \bridgek \qquad {\cal V}_{k\over 2}....
\eqno(44)
$$

In order to study how the quantum group acts on the Hilbert
space, we investigate the transformation of ${\cal U}$
under the quantum group symmetry generated by (38).
Using (39) and (43), we can write down the following commutators
$$\eqalign{
& t^{\pm S_3} {\cal U}_{ab} = {\cal U}_{ab} t^{\pm S_3}
t^{\pm a/2}, \qquad \qquad \forall a,b = \pm 1;\cr
& [t^{-S_3} S_\pm, {\cal U}] = -t^{-2S_3 \pm \half} \sigma_\mp
{\cal U}.  \cr}
\eqno(45)
$$
The second equation in (45) can also be written as
$$
S_\pm ({\cal U}_{+b},{\cal U}_{-b}) = ({\cal U}_{+b},{\cal U}_{-b})
\left( t^{\half \sigma_3} S_\pm - t^{\mp 1} \sigma_\pm  t^{-S_3}
\right) .
\eqno(46)
$$
This implies that the transformation law of ${\cal U}$ under
$U_t(SL(2))$ symmetry can be formulated in terms of the
coproducts; namely, for every element $\varrho \in U_t(SL(2))$,
$\varrho$ acts on ${\cal U}$ in the representation of
spin $\half$ according to
$$
\varrho  {\cal U} ={\cal U} ({\Pi}^{[\half]} \o id ) \Delta (\varrho).
\eqno(47)
$$
We have denoted ${\Pi}^{[\half]}(\varrho)$ to be the
spin $\half$ representation of $\varrho$.
These cocycles are determined by the ones for the
generators of $U_t(SL(2))$,
$$\eqalign{
& \Delta (t^{S_3}) = t^{S_3} \o t^{S_3}, \cr
& \Delta (S_\pm) = t^{S_3} \o S_\pm + S_\pm \o t^{-S_3}. \cr}
\eqno(48)
$$

It is also possible to study the commutation relation of the
deformed Wigner operator ${\cal U}$ with itself. It turns out that
it obeys the following exchange relation,
$$
R {\cal U}_1 {\cal U}_2 = {\cal U}_2 {\cal U}_1 B(p),
\eqno(49)
$$
where $B(p=\ell+\half)$ gives the Racah $\{ 6j \}$-matrix\ref{\KR}
for $U_t(SL(2))$,
$$
B(p)=\pmatrix{
t^{\half} & 0 & 0 & 0 \cr
0 & t^{-\half} \sqrt{[2p+1][2p-1]\over[2p]^2} & - t^{-\half-2p}
{1\over[2p]} &  0 \cr
0 & t^{-\half+2p} {1\over [2p]} & t^{-\half}
\sqrt{[2p+1][2p-1]\over[2p]^2} & 0 \cr
0 & 0 & 0 & t^{\half} \cr},
\eqno(50)
$$
and $R$ is the constant R-matrix for $U_t(SL(2))$,
$$
R=t^{-\half} \pmatrix{
t & 0 & 0 & 0 \cr 0 & 1 & 0 & 0 \cr
0 & t - t^{-1} & 1 & 0 \cr 0 & 0 & 0 & t \cr}.
\eqno(51)
$$
As it turns out, Eq. (49) gives the operator identity for the
IRF-Vertex transformation discussed in [{\RESH}].

In this realisation of $U_t(SL(2))$,
we do not restrict the value of the parameter $t$.
In the limit when $t= 1$, we recover the results of
the undeformed model in the previous section.
For the interest of the chiral WZW models,
we need to consider when the
deformation parameter $t$ is a root of the unity, i.e.
$t=e^{i{\pi\over k+2}}$ for some positive integer $k$.
In this case, we have only a finite number of
integrable representations which corresponds to
$\ell=0,\half,1,...k/2$. Although the braiding matrix
in (50) becomes singular when $p=0$ or $p= (k+2)/2$,
i.e. $\ell=-{1\over 2}$ or $\ell=(k+1)/2$,
these singularities are not harmful when we
consider only the integrable representations.

\sect{Symplectic Structure and the Lagrangian formulation
for $U_t(SL(2))$}

We have given a simple realisation of the quantum group
generators and the Wigner operator ${\cal U}$ in terms of two
pairs of harmonic oscillators. In order to understand the
geometrical meaning of this quantum theory and the origin of the
deformation, one would like to have a Lagrangian formulation for
its classical theory. It seems very plausible that
${\cal U}$ can be taken as the defining variable (particle)
which transform covariantly under $U_t(SL(2))$.
Indeed, by taking the classical limits of the quantum relations
in the previous section, we obtain a set of Poisson brackets
given by the following closed two-form in terms of ${\cal U}$,
$$\eqalign{
\omega &= i \delta q \delta p + i \delta A \delta D, \cr
       &= {i\over 4\gamma} Tr \{  i 4\gamma \delta p \sigma_3
{\cal U}^{-1} \delta{\cal U}  + e^{-i2\gamma p \sigma_3}
{\cal U}^{-1}\delta {\cal U}
e^{i2\gamma p \sigma_3} {\cal U}^{-1}\delta {\cal U} \cr
    & \qquad + {1\over 4} (Z_+^{-1}\delta Z_+ + {\hat w} (Z_+^{-1}\delta Z_+)
{\hat w}^{-1})(\delta Z_- Z_-^{-1} + {\hat w}
(\delta Z_- Z_-^{-1}){\hat w}^{-1}) \} \cr}
\eqno(52)
$$
where ${\hat w}=\left( \matrix{0&1\cr 1&0\cr} \right)$
is the Weyl reflection and $Z_+$ and $Z_-$ are elements in the
Borel subgroups $B_+$ and $B_-$ defined in
$$
Z=Z_+ Z_- \equiv {\cal U} e^{i2\gamma p \sigma_3} {\cal U}^{-1}.
\eqno(53)
$$
Although $Z_\pm$ are defined up to an element $h$ in the maximal
torus, i.e. $Z_+ \rightarrow Z_+ h$, and $Z_- \rightarrow h^{-1} Z_-$,
the symplectic form in (52) is not affected by this re-definition.
$\gamma$ is related to the deformation parameter according to
$t=e^{i\gamma}$.

A similar symplectic form had been discussed by
Alekseev and Malkin\ref{{\AM}} in a more general context.
Their approach is to deform the
Kirillov symplectic form for Lie-Poisson groups.
It is plausible that our result provides an explicit
example of their approach.
Once the symplectic form is determined,
we can choose the classical Hamiltonian $H_0$ to be
quadratic Casimir given in (40) and write down
the Lagrangian for the interacting particle
with $U_t (SL(2))$ symmetry as
$$
L= Tr \left\{ { \Pi}({\cal U}) \partial_\tau {\cal U} - H_0 \right\},
\eqno(54)
$$
where $\Pi({\cal U})$ is the conjugate momentum of ${\cal U}$ as in
the symplectic form $\omega = Tr \{ \delta {\cal U}\delta \Pi \}$.
However, it is difficult to obtain a simple way of writing
the Lagrangian in terms of ${\cal U}$ other than
a perturbative expression in powers of $\ln t$.
We hope to consider this further in the near
future.

\sect{Acknowledgements}

We would like to thank David Olive for helpful discussions
in the early stage of this work, Fedor Smirnov for
suggesting the connection of our work to
the IRF-Vertex relation and Shahn Majid for the
co-product formulation of the quantum group
transformation laws. Part of this work was done during M. Chu's
visit at the Aspen Center for Physics in August 1993.
This work is supported by the Science
and Engineering Research Council under the grant GR/H57929.

\vskip 12pt
\leftline{\bf References}

\item{ [1]} V. G. Drinfeld, ``Proceedings of International Congress
of Mathematics, Berkeley", American Mathematical Society, 1986.

\item{ [2]} M. Chu, P. Goddard, I. Halliday, D. Olive, A. Schwimmer,
Phys. Lett. {\bf B266} (1991) 71.

\item{ [3]} L.D. Faddeev, Commun. Math. Phys. {\bf 132} (1990) 131.

\item{ [4]} A. Yu. Alekseev, L.D. Faddeev,
Commun. Math. Phys. {\bf 141} (1991) 413.

\item{ [5]} G. J. Zuckerman, {\it Action principles and global geometry},
in ``Mathematical Aspects of String Theory'',
edited by S.T. Yau, World Scientific, Singapore, 1987.

\item{ [6]} A. Messiah, Quantum Mechanics, Vol. II, North Holland,
Amsterdam, 1970.

\item{ [{\RESH}]} N. Yu. Reshetikhin,``Quantised Universal
Enveloping Algebras, The Yang-Baxter Equation and Invariants
of Links, I, II", LOMI preprint E-4-87.

\item{ [{\KR}]} A.N. Kirillov, N. Yu. Reshetikhin, ``Infinite-dimensional
Lie algebras and groups" edited by V. G. Kac, World
Scientific, Singapore, (1989) pp 285.

\item{ [{\AM}]} A. Yu. Alekseev, A. Z. Malkin, Commun. Math. Phys.
{\bf 162} (1994) 147-173.

\end

In principle, given the symplectic form in ${\cal U}$,
one can write down the Lagrangian in terms of ${\cal U}$.
The idea is to write the symplectic form as
$$
\omega = Tr \left\{ \delta {\cal U} \delta { \Pi} \right\}
\eqno(61)
$$
and choose the quadratic Casimir to be the Hamiltonian,
$$
H ({\cal U},{\Pi}) =S_+ S_- + [S_3]_t [S_3+1]_t.
\eqno(62)
$$

In principle, given the symplectic form in ${\cal U}$,
one can write down the Lagrangian in terms of ${\cal U}$.
The idea is to write the symplectic form as
$$
\omega = Tr \left\{ \delta {\cal U} \delta { \Pi} \right\}
\eqno(61)
$$
and choose the quadratic Casimir to be the Hamiltonian,
$$
H ({\cal U},{\Pi}) =S_+ S_- + [S_3]_t [S_3+1]_t.
\eqno(62)
$$